\documentstyle[epsf]{mn}
\newcommand{\un}[2]{\mbox{\rm\thinspace #1$^{#2}$}}

\newcommand{\be}[1]{\begin{equation}\label{#1}}
\newcommand{\ee}{\end{equation}}

\newcommand{\gsim}{\mathrel{\hbox{\rlap{\lower.55ex \hbox {$\sim$}}
                   \kern-.3em \raise.4ex \hbox{$>$}}}}
\newcommand{\lsim}{\mathrel{\hbox{\rlap{\lower.55ex \hbox {$\sim$}}
                   \kern-.3em \raise.4ex \hbox{$<$}}}}
\newcommand{\msun}{\mbox{M$_\odot$}}

\newcommand{\sub}[1]{_{\rm #1}}

\newlength{\ffh}

\title[Gamma-ray bursts from stellar remnants]
      {Gamma-ray bursts from stellar remnants: probing the Universe
           at high redshift}

\author[Wijers et al.]
       {Ralph A.M.J. Wijers\thanks{Email: ramjw@ast.cam.ac.uk}, 
        Joshua S. Bloom\thanks{present address: Caltech, 105-24, 
	Pasadena, CA 91125, USA}, Jasjeet S. Bagla and Priyamvada Natarajan\\
        Institute of Astronomy, University of Cambridge, Madingley Road,
        Cambridge CB3 0HA, U.K.}

\date{Submitted to MNRAS}

\pagerange{\pageref{firstpage}--\pageref{lastpage}}
\pubyear{1997}

\begin{document}

\label{firstpage}

\maketitle

\begin{abstract}
A gamma-ray burst (GRB) releases an amount of energy similar to that of a
supernova explosion, which combined with its rapid variability suggests an
origin related to neutron stars or black holes. Since these compact stellar
remnants form from the most massive stars not long after their birth,
gamma-ray bursts should trace the star formation rate in the Universe;  we
show that the GRB flux distribution is consistent with this.  Because of
the strong evolution of the star formation rate with redshift, it follows
that the dimmest known bursts have $z\sim6$, much above the value usually
quoted and beyond the most distant quasars.  This explains the absence of
bright galaxies in well-studied gamma-ray burst error boxes. The increased
distances imply a peak luminosity of 
$8.3\times10^{51}\un{erg}{}\un{s}{-1}$
and a rate density of 0.025 per million years per galaxy. These values are
20 times higher and 150 times lower, respectively, than are implied by fits
with non-evolving GRB rates.  This means that GRBs are either caused by a
much rarer phenomenon than mergers of binary neutron stars, or their
gamma-ray emission is often invisible to us due to beaming. Precise
burst locations from optical transients will discriminate between the
various models for GRBs from stellar deaths, because the distance between
progenitor birth place and burst varies greatly among them.  The dimmest
GRBs are then the most distant known objects, and may probe the Universe at an
age when the first stars were forming.
\end{abstract}


\section{Introduction}

The ability of the Italian-Dutch BeppoSAX satellite \cite{psb:95} to
accurately locate gamma-ray bursts (GRBs) with its Wide Field
Cameras \cite{jhzb:95} has led to the discovery of two optical
transients, associated with GRB\,970228 \cite{pggks:97} and
GRB\,970508 \cite{bond:97}.  The detection of redshifted absorption
lines in the optical transient associated with
GRB\,970508 \cite{mdska:97} has established that it lies at
cosmological distance, and here we assume they all do. 

A typical dim burst ($1\un{ph}{}\un{cm}{-2}\un{s}{-1}$ lasting 10\,s) at
$z=1$ releases $3.5\times10^{49}\un{erg}{}\un{sr}{-1}$ in gamma rays, hence
the engine behind it must provide about
$4.4\times10^{52}f\sub{b}/\epsilon_{-2}\un{erg}{}$. (Where needed, we 
assume $H_0=70\un{km}{}\un{s}{-1}\un{Mpc}{-1}$ and $\Omega_0=1$.)
Here $f\sub{b}$ is the
fraction of the sky illuminated by the gamma-ray emission, and the
efficiency of conversion of the initially available energy into gamma rays
is $\epsilon_{-2}$ percent. This is the natural energy scale of supernova
explosions, in which of order $10^{53}$\,ergs is released suddenly from the
rest mass energy of a solar mass of material. The bulk is carried away in
neutrinos, and about 1\% becomes kinetic energy of the ejecta.
The proposed GRB models related to end stages of massive stars
are (i) merger of two
neutron stars \cite{paczy:86,gdn:87,elps:89}; (ii) merger of a neutron
star and a black hole \cite{paczy:91,mhim:93}; (iii)
`failed supernova': the collapse of a massive star to a black hole
surrounded by a dense torus of material that might result in a
relativistic jet \cite{woosl2:93}; (iv) a `hypernova': the collapse of a
rapidly rotating massive star in a binary \cite{paczy:97};
(v) collapse of a Chandrasekhar-mass
white dwarf \cite{usov:92}. Whether these are efficient enough at
converting a fraction of the available energy to kinetic energy and then
eventually to gamma rays (see below) is an open question, and the major
unsolved issue in this class of burst models. In this paper we assume that
somehow a variety of such a model manages this. The important point is
that they all arise from massive stars which evolve into remnants within 
about 100\,Myr. The binary mergers then usually take place within about
100\,Myr of remnant formation \cite{ps:96}, as does the white-dwarf
collapse, because the favoured route has a high mass transfer rate
\cite{hbnr:92}.  Since the expansion age of the Universe is already
1\,Gyr at $z=4.4$, it is safe to neglect the delay between (binary) stellar
birth and the GRB it eventually yields in the present
context.  {\it The gamma-ray burst rate therefore traces the massive
star formation rate.\/} The star formation rate  as a function of redshift
has recently been studied extensively, and is determined observationally
with some confidence \cite{llhc:96,mfdgs:96,madau:96}: the luminosity
density in the rest frame $U$ and $B$ band is combined with an IMF to
deduce the star formation rate. The assumption of an IMF introduces an
uncertainty in the deduced total star formation rate, but the basic data
($U$ and $B$ light density) are dominated by massive stars. Since GRBs come
from massive stars, they may trace the UV light density in the Universe
better than the total star formation rate, and our results are therefore
less sensitive to the assumed IMF.
A further potential source of uncertainty is dust extinction,
which would cause a relative
underestimate of the high-redshift star formation rate.

Recent interpretations of the afterglows \cite{mr:97,wrm:97,waxma:97}
support the notion that the energy release is initially in the
form of an ultrarelativistic explosion or `fireball' \cite{cr:78,goodm:86}
whose energy is largely converted to a blast of gamma rays via hydrodynamic
collisions within it \cite{px:94,rm:94} or with the ambient medium
\cite{rm:92}. Since the kinetic energy comes from a fairly standardised
event, it is likely that the gamma-ray luminosity distribution of
bursts is not too wide, so we shall treat them as standard candles. With
the fact that GRB trace star formation, this has the important testable
consequence that the redshift dependence of the GRB rate has
no free parameters. Only two normalisations need to be fitted, namely the
local GRB rate density $\rho_0$ and the standard-candle 
30--2000\,keV luminosity $L_0$.

\section{Results}

\begin{figure}
   \epsfxsize=\columnwidth\epsfbox{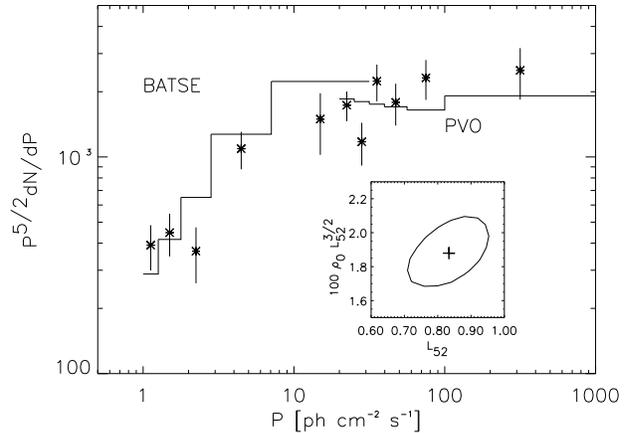}
   \caption[]{The flux distribution of GRBs from PVO
              and BATSE and the best-fit model proportional to the star
              formation rate. The $y$ axis is $P^{5/2}$ times the rate,
              which is period-independent at high fluxes, to emphasise this
              asymptotic behaviour and the turnover at low fluxes.  The
              inset shows the 1-$\sigma$ confidence region of the fitted
              parameters.  The apparent mismatch at the PVO-BATSE transition
              is just due to the differences in bin sizes.
              \label{fi:fit} 
              } 
\end{figure}

For a standard cosmology ($\Omega_0=1$, $\Lambda=0$), the predicted
number of standard-candle bursts in some flux range (${\rm P_1~to~P_2}$) is
\begin{equation}
\Delta N({\rm P_1~to~P_2}) =  4\pi \int_{R(P_1)}^{R(P_2)} 
k_\rho\rho(z) r^2 dr~~,
\end{equation} 
where $\rho(z)$ is the observed star formation rate and $k_\rho$ the constant of
proportionality. We follow the method
of Fenimore and Bloom (1995) to account for the influence of the
diversity of spectral shapes of bursts on the observed flux distribution 
(similar to $K$ corrections in optical photometry). The fit is done by $\chi^2$ 
minimisation for the same
11 flux bins of combined PVO and BATSE data used by Fenimore and Bloom (1995).
The best-fit model of this type to the GRB flux distribution is shown
in Fig.~\ref{fi:fit}. Note that the fit was done only for 
$P>1\un{ph}{}\un{cm}{-2}\un{s}{-1}$, for which the BATSE catalogue is 99\%
complete.

The fit gives $L_0=8.3^{+0.9}_{-1.5}\times10^{51}\un{erg}{}\un{s}{-1}$ and
$\rho(z=0)\equiv\rho_0=0.14\pm0.02\un{Gpc}{-3}\un{yr}{-1}$. Assuming a
local galaxy number density of 0.0048$\un{Mpc}{-3}$ \cite{lpem:92}
this density translates
into a rate of 0.025\,GEM (Galactic Events per Myr). The median redshift of
bursts with $P=1\un{ph}{}\un{cm}{-2}\un{s}{-1}$ is 2.6. The fit is
just acceptable, with $\chi^2=17.3$ for 9 d.o.f. 
If we omit the two lowest-flux bins, in which the star formation rate is
most uncertain, the fitted parameters
hardly change but the fit quality improves somewhat to
$\chi^2/{\rm d.o.f.}=11.7/7$. As noted earlier, inclusion of dust extinction
would increase the inferred star formation rate at high redshift, which
in turn would improve the fit. However, the magnitude of this correction
is quite uncertain.

For comparison, we also fitted the same data with a non-evolving rate
density.  In that case, we recover the previously known result that
$L_0=0.44\times10^{51}\un{erg}{}\un{s}{-1}$ and $\rho_0=3.7$\,GEM
\cite{fb:95}. The redshift at $P=1\un{ph}{}\un{cm}{-2}\un{s}{-1}$ is then
0.68, and $\chi^2/{\rm d.o.f.}=9.1/9$. This means that our assumption
about the evolution of the GRB rate has quite drastic consequences: it
increases the GRB luminosity by a factor 19, and the local rate is
decreased by a factor 150.  This large factor is a combination of the
distance increase due to the larger $L_0$ and the fact that the local
density in evolving models is much lower than the mean, because the star
formation rate in the Universe was much higher at $z=1$. Various indirect
methods, such as time dilation \cite{nbnsk:95,fb:95} and the change of
break energy with flux \cite{mppbp:95}, have been used to statistically
derive the ratio of redshifts between bright and dim bursts. The result was
found to disagree with the low redshifts implied by fits to the flux
distribution that assumed no evolution of the rate density 
\cite{fb:95,brain:97}. For our new value of $L_0$, the predicted time dilation
factor is 1.9 \cite{fb:95}, consistent with the measured value
\cite{nbnsk:95}.  Note that the slope of the cumulative flux distribution
of $-1.5$ at moderate fluxes is not due to Euclidean geometry: it is a
conspiracy between the curvature of space which tends to give a flatter
slope and the strong evolution which gives a steeper one.
Direct support of the significant redshift of bursts on the `Euclidean' part
of the flux distribution comes from the work of Dezalay et~al. (1997a,b).
From a detected hardness-intensity correlation in bright bursts seen by
both ULYSSES and PHEBUS they infer $z\simeq3$ for bursts that roughly
correspond to $P=2\un{ph}{}\un{cm}{-2}\un{s}{-1}$ in BATSE terms. This is
even slightly higher than our range of $z=1.5-2.5$ at that flux level.

\section{Implications}

The twenty-fold increase in luminosity of the bursts has important
implications. First, the total energy released in gamma rays in a 10-s
burst goes up to $6.6\times10^{51}\un{erg}{}\un{sr}{-1}$, requiring an
initial supply of energy of
$8.3\times10^{54}f\sub{b}/\epsilon_{-2}\un{erg}{}$, which among the
above mechanisms only the hypernova \cite{paczy:97}
can provide if the emission is isotropic. We
therefore conclude that gamma-ray beams of bursts probably
illuminate no more than a few percent of the sky, hence
most gamma-ray bursters escape detection in gamma rays.  Since all the
models of interest entail the collapse of an already rotating system and a
non-vanishing angular momentum implies cylindrical symmetry, such beaming
is quite plausible in the context of these models. The higher
amount of energy alleviates the baryon pollution problem: for the outflow to
reach a Lorentz factor above 100, as required to produce gamma rays, 
it should contain at most $10^{-5}\msun$ of baryons, which is not easy
to get; but now we can allow twenty times more.

The classical GRB rate is 1\,GEM, but we now find a rate of
only 0.025\,GEM. Since the events could well be beamed, this does not
necessarily exclude neutron star mergers as their source. The theoretical
rate of NS-NS mergers is about 10\,GEM \cite{ps:96}, implying
$f\sub{b}=0.3\%$ if all such mergers produce a GRB. (See also
\cite{lpp:97,plp:97}, where similar conclusions are drawn using theoretical
rather than observed star formation rates.) But it does mean that rarer
types of event merit consideration as well. NS-BH mergers are probably
about ten times rarer than NS-NS mergers, so they could be significantly
beamed and still cause the observed GRB rate. The formation rate of
super-soft X-ray sources is about 20 GEM \cite{hbnr:92}, but since it is not
known what fraction, if any, of these lead to the accretion-induced
collapse of a white dwarf and a possible GRB therefrom
\cite{usov:92} we cannot calculate a rate for this burst model.

The increased distance scale also removes the `no-host problem' for
GRBs: deep searches of GRB error boxes
\cite{vhj:95,fehkl:93,schl:97} have been used to set limits on the
absolute brightness of host galaxies of 3.5 to 5.5 {\it mag\/} fainter than
$L_\star$ \cite{schl:97}, suggesting that GRB do not come from galaxies.
But since these estimates depend on $L_0$ and now have to be adjusted by 
3.2 {\it mag\/} just from the increased
distance, and by another 1--2 {\it mag}, depending on galaxy type, due to
increased $K$ corrections \cite{lthcl:95}.  This changes the limits
on host galaxy luminosities to between 1.5 {\it mag\/} above and 1.5 
{\it mag\/} below
$L_\star$, so they are no longer inconsistent with the assumption that GRB
are in normal galaxies. 

Using our fit for $L_0$ and the known gamma-ray spectra we can derive
redshifts for the two GRB with detected optical afterglows. We find $z=2$
for GRB\,970228, consistent with the magnitudes of candidate hosts
\cite{btw:97}. But $z=3.7$ for GRB\,970508, which exceeds the maximum
redshift of 2.3 allowed by the observed spectrum \cite{mdska:97} and shows that
this burst must have been somewhat less luminous. This means that its host
is 2.5 to 5 magnitudes fainter than $L_\star$ \cite{nbsjt:97}, and that the
luminosity distribution of GRB must have some width. 
Fortunately, the shape of the GRB flux distribution is fairly 
insensitive to a modest broadening of the luminosity function (e.g.\
Ulmer, Wijers \& Fenimore 1995) so this should not significantly influence
our conclusions.

\begin{figure}
   \epsfxsize=\columnwidth\epsfbox{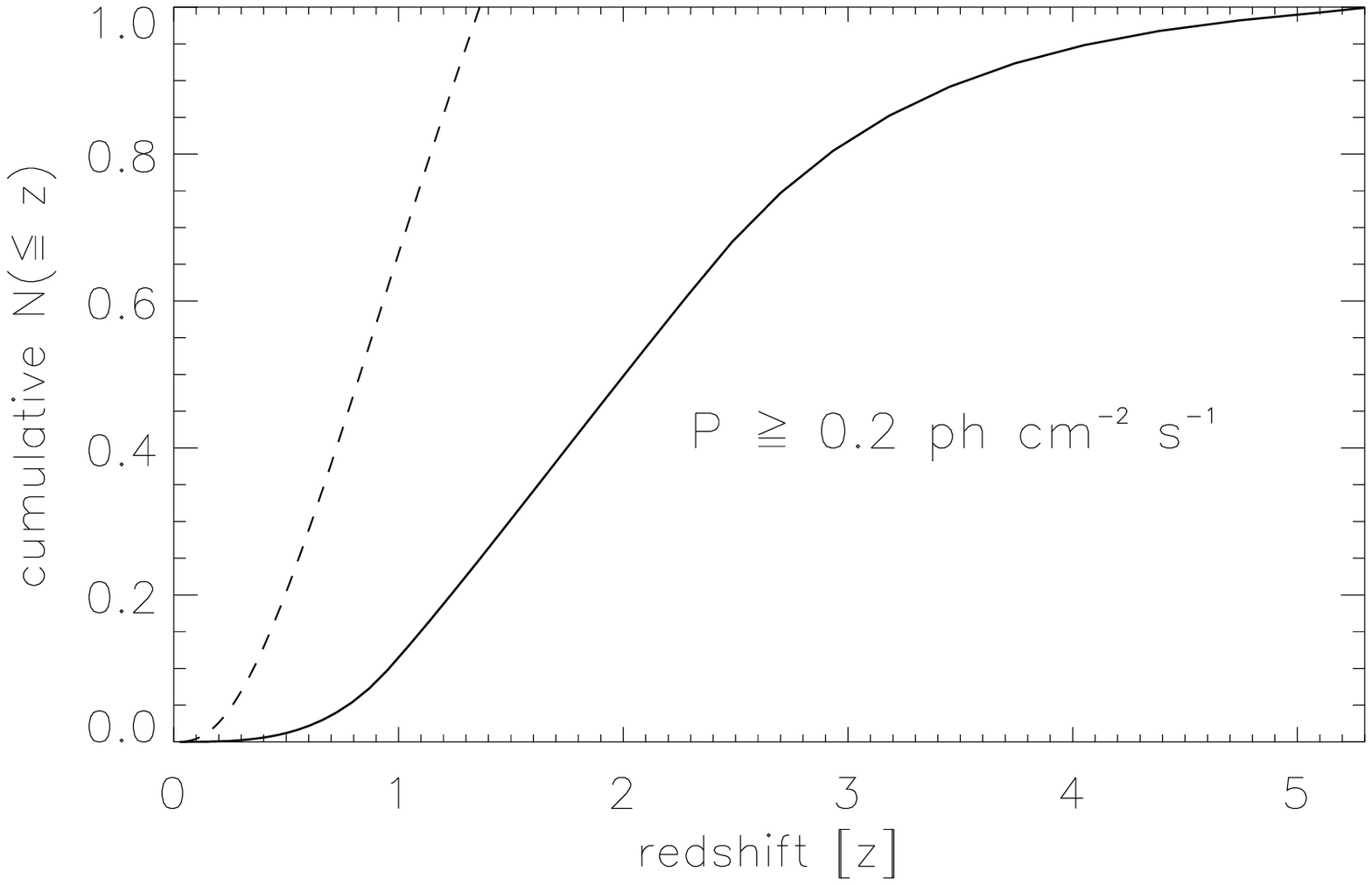}

   \epsfxsize=\columnwidth\epsfbox{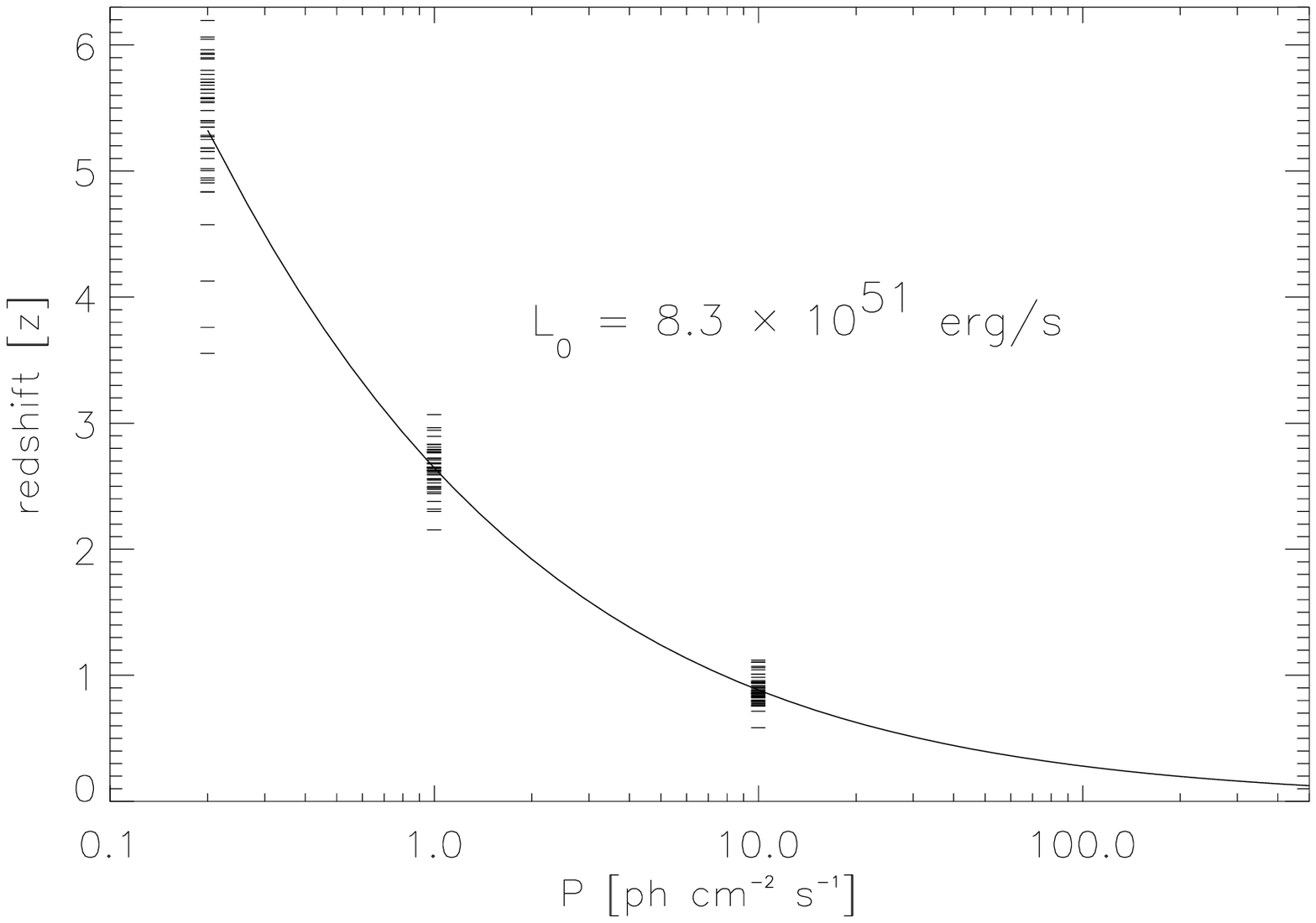}
   \caption[]{(top) The fraction of GRBs below redshift $z$
              according to the best-fit evolving model, for bursts with the
              median spectral shape down to
	      $0.2\un{ph}{}\un{cm}{-2}\un{s}{-1}$ in the evolving (solid)
	      and non-evolving (dashed) case. (bottom) The redshift as
	      a function of flux for the evolving model fit. The solid curve
	      gives the average over spectral shapes. For three
	      flux values, the individual redshifts for each of 48 measured
	      spectra \cite{bmfsp:93} are shown to indicate the 
	      considerable variation due to spectral shape.
              \label{fi:nz}
             }
\end{figure}

The best-fit redshift distribution of GRBs
is shown in Fig.~\ref{fi:nz}. The median redshift is similar to that of 
quasars; only 10\% of bursts have $z<1$, and 5\% are
beyond redshift 4.  The dim end of the distribution is at $z=5.3$ for
the average spectrum, but goes up to $z=6.2$ due to varying spectral shape.

Sahu et~al.\ (1997) briefly discuss the possibility of GRBs following
the star formation rate. Whilst they do not account for the variety
of spectral shapes and do not perform a formal fit, they conclude from
visual inspection of their graphs that $L_\gamma=10^{51}\un{erg}{}\un{s}{-1}$.
They interpret this as meaning that the accepted standard luminosity 
fits the data. However, since they count $L_\gamma$ from 100 to 500\,keV
instead of the range 30--2000\,keV used in this and other works, their
$L_\gamma$ implies $L_0\simeq3L_\gamma$. This is closer to our new value
than to the no-evolution result, and in reasonable agreement with our
$L_0$ given the differences between the methods.

Totani (1997) tried different power-law spectral shapes, but did not allow
for variation between bursts and only considered the case of NS-NS mergers.
He did study the issue of NS-NS merger times in more detail and found that
the tail of late mergers can flatten the flux distribution between
redshifts 0 and 1: if a substantial fraction of mergers have a long delay,
then because 20 times more binaries formed at $z=1$ their contribution to
the present merger rate could exceed that due to current star formation.
Whilst the bulk of NS-NS binaries merge within 100\,Myr according to all
studies, the fraction merging after more than 5\,Gyr varies greatly.
Tutukov \& Yungelson (1994) do find a large fraction of delayed mergers,
whereas in the study of Lipunov et~al.\ (1995) it is negligible. Using this
long delay, he finds a redshift of 2--2.5, for bursts with
$P=0.4\un{ph}{}\un{cm}{-2}\un{s}{-1}$, where we get a median of 3.8.
Whether this effect is indeed important can only be decided when better
estimates of the merger time distribution become available. If GRB are not
due to NS-NS mergers (we present some evidence for this below) then the
long delay times are not an issue in any case, of course.

\section{Observational tests}

An important difference between the various compact stellar remnant models
is the distance that a gamma-ray burster travels between where it was born
as a massive (binary) star and where it produces the burst. A direct
supernova origin \cite{woosl2:93} or hypernova \cite{paczy:97}
would occur in short-lived objects with low space velocity, 
which would therefore
still be in the star forming regions where they were born. In this case,
an optical counterpart to a GRB should always be embedded in a
galaxy or star-forming region.  A NS-NS or NS-BH
merger occurs in a system that has obtained a moderate to
high (100--300$\un{km}{}\un{s}{-1}$) space velocity from the two supernova
explosions that have taken place in the binary some 100\,Myr before the
merger. This means that such GRB should often occur up to 30\,kpc away from any
star forming region (corresponding to 6'' at $z=3$) depending on whether
the host galaxy has a deep enough potential to hold it. 
The optical counterpart
to GRB\,970228 is embedded in an extended object. That of GRB\,970508
is at least 25\,kpc away from any host so far detected \cite{nbsjt:97},
but the [OII] emission line seen in its spectrum (Metzger et~al.\ 1997)
suggests that it lies in a star-forming region. The recent detection
of a large absorption column in the X-ray spectrum of GRB\,970828 
(Murakami 1997) suggests that it, too, may lie close to a star-forming region.
There may thus be some tentative evidence favouring progenitors with
low space velocities and very short delays between formation and burst.

Another consequence of beamed gamma-ray emission in the context of blast
wave models  is that the optical afterglow can come from less relativistic
material which has a greater opening angle \cite{wrm:97,paczy:97}.
Consequently, the population of
bursters that we can see only by their optical afterglow could be many
times larger than that of GRBs. A limit to this is set by
high-redshift supernova searches, which have not found any GRB afterglows
\cite{rhoad:97}.
Since one of them \cite{phdgg:96} has now surveyed close to 10 square
degree years, a rate of afterglows in excess of 0.3 per square degree per
year is unlikely. With a GRB rate of 0.01/sq.deg./yr, this implies that
$f\sub{b,optical}/f\sub{b,\gamma}\lsim30$. Since $f\sub{b,\gamma}\lsim0.03$
from energy constraints, even the optical afterglows may have to be beamed.

Since with the evolving rate density we sample most of the star forming
Universe, we predict that more sensitive instruments than BATSE would find few
bursts with $P<0.2\un{ph}{}\un{cm}{-2}\un{s}{-1}$, unlike in the
non-evolving case (see Fig.~\ref{fi:nz}). Kommers et~al.\ (1997)
recently studied untriggered bursts below the BATSE threshold and 
deduced that the flux distribution flattens considerably, perhaps
supporting the paucity of faint bursts.
At very low fluxes, there may again be increase due to the
first episode of star formation 
in the early Universe associated with the first metal production
\cite{mr2:97} or with the formation of the first stars in elliptical
galaxies \cite{plp:97}.

With a higher mean redshift than quasars, GRB should be lensed at least as
often as quasars, of which 0.5\% are multiply imaged.  Therefore, the dim
end of the BATSE catalogue may already contain a few examples.  From the
absence of lensed bursts among bright BATSE bursts Marani et~al.\ (1997)
deduce that bursts with $P>1\un{ph}{}\un{cm}{-2}\un{s}{-1}$ should have
$z<3$; our fit is consistent with this limit.

\section{Conclusions}

Our results depend only on two key features of the star formation rate:
rapid evolution up to $z\approx 1$ and a more gentle variation beyond that
up to $z \approx 2.5$.  Although little is known about the star formation
rate between these redshifts, many indirect arguments suggest that the star
formation rate did not evolve strongly between $z=1$ and $z=2.5$.
Uncertainties are introduced by our lack of knowledge of the correction due
to dust extinction and to a lesser extent by having to assume an IMF.
However this does not change the qualitative nature of the results
presented here.

Furthermore, the distance corrections due to
the wide range in spectral shapes \cite{fb:95} that are the 
result of assuming that GRB are standard candles in the 30--2000\,keV
(rest frame) range are very important, and neglect of the spectral shape
variation can lead to considerably different results.
In view of this, and of the fact that we have not explored a range of 
parameters for the geometry of the Universe, our results obviously
have a somewhat exploratory character. Other observational tests, such as
the determination of redshifts from afterglow spectra and constraints from
the lensing rate of GRBs, will provide additional constraints on parameters.

The logical consequence of assuming that GRB are related to remnants of the
most massive stars in any of the ways hitherto proposed is that the GRB
rate is proportional to the formation rate of massive stars in the Universe
(with the possible exception of NS-NS mergers if those simulations which
indicate significant fractions of late mergers are correct).  We show that
this assumption is consistent with the GRB flux distribution.  Compared to
previous, non-evolving, models of cosmological bursts we find a twenty-fold
increase of the required GRB luminosity, which suggests that
the gamma-ray emission is significantly beamed in order that the
emitted energy can be supplied by merger/collapse models. The redshifts of
GRBs are also greatly increased, and the very dimmest known ones are at
$z\gsim6$, beyond the farthest known quasars. This makes them the most
distant known objects, and their optical counterparts very valuable probes
of the early evolution of stars and interstellar gas.

%%%%%%%%%%%%%%%%%%%%%%%%%%%%%%%%%%%%%%%%%%%%%
%  tail stuff: bibliography
%%%%%%%%%%%%%%%%%%%%%%%%%%%%%%%%%%%%%%%%%%%%%
%\bibliographystyle{unsrt}
%\bibliography{xmymonic,x65,x70,x75,x80,x85,x90,x95,x00,xnew,zstargrb}

\section*{ACKNOWLEDGEMENTS}

We thank S. McGaugh, R. McMahon, J. Miralda, M. Rees, and J. Wambsganss
for helpful discussions and comments.

\label{lastpage}

\end{document}